\documentclass[10pt,conference,english]{IEEEtran}
\IEEEoverridecommandlockouts
\usepackage[centertags]{amsmath}
\usepackage{amsfonts}
\usepackage{amssymb}
\usepackage{epsfig}
\usepackage{amsmath,  amssymb}
\usepackage{graphicx}
\usepackage[centertags]{amsmath}
\usepackage{clrscode}
\usepackage{cite}
\usepackage{fullpage}
\usepackage{amsmath,amssymb,mathrsfs,pseudocode}
\usepackage{color}
\usepackage[normalem]{ulem}
\usepackage{psfrag}
\usepackage{url}
\usepackage{babel}
\usepackage{authblk}
\usepackage{mathtools}
\usepackage[linesnumbered,ruled]{algorithm2e}
\usepackage[left=54pt, right=54pt, top=51pt, bottom=58pt]{geometry}
\usepackage{xspace}% essential for newcommand endings
\usepackage{multicol}
\usepackage{booktabs} % nice table objects
\usepackage{multirow}
\usepackage[x11names,dvipsnames,table]{xcolor}

\usepackage{subfigure}

\usepackage{stackrel}
\usepackage{extarrows}

\newtheorem{theorem}{Theorem}%[section]
\newtheorem{lemma}{Lemma}

\newtheorem{proposition}{Proposition}%[section]
\newtheorem{example}{Example}%[section]
\newtheorem{definition}{Definition}%[section]
\newtheorem{corollary}{Corollary}%[section]
\newtheorem{remark}{Remark}%[section]

\usepackage{xspace}
\usepackage{bbm}
\input{mathlig}

\newcommand{\muspace}{\mspace{1mu}}

\DeclareRobustCommand{\scond}{\mathchoice{\muspace\vert\muspace}{\vert}{\vert}{\vert}}
\mathlig{|}{\scond}

\DeclareRobustCommand{\discint}{\mathchoice{\mspace{-1.5mu}:\mspace{-1.5mu}}{\mspace{-1.5mu}:\mspace{-1.5mu}}{:}{:}}
\mathlig{::}{\discint}

\newcommand{\Ec}{\mathcal{E}}

\newcommand{\Hc}{\mathcal{H}}
\newcommand{\Ic}{\mathcal{I}}

\newcommand{\Kc}{\mathcal{K}}
\newcommand{\Lc}{\mathcal{L}}
\newcommand{\Mc}{\mathcal{M}}

\newcommand{\Rc}{\mathcal{R}}
\newcommand{\Sc}{\mathcal{S}}
\newcommand{\Tc}{\mathcal{T}}

\newcommand{\Vc}{\mathcal{V}}

\newcommand{\Xc}{\mathcal{X}}
\newcommand{\Yc}{\mathcal{Y}}
\newcommand{\Zc}{\mathcal{Z}}

\newcommand{\Gcal}{\mathcal{G}}

\newcommand{\Kcal}{\mathcal{K}}

\newcommand{\Pcal}{\mathcal{P}}

\newcommand{\Xcal}{\mathcal{X}}

\newcommand{\Zcal}{\mathcal{Z}}

\newcommand{\Av}{{\bf A}}

\newcommand{\Xv}{{X}}

\newcommand{\Wv}{{\bf W}}

\newcommand{\dv}{{\bf d}}

\newcommand{\gv}{{\bf g}}

\newcommand{\xv}{{x}}

\newcommand{\Xh}{{\hat{X}}}

\newcommand{\Yt}{{\tilde{Y}}}

\newcommand{\xt}{{\tilde{x}}}

\def\d{\delta}
\def\e{\epsilon}

\def\textiid{i.i.d.\@\xspace}
\newcommand\iid{\ifmmode\text{ i.i.d. } \else \textiid \fi}

\def\clap#1{\hbox to 0pt{\hss#1\hss}}
\def\mathclap{\mathpalette\mathclapinternal}
\def\mathclapinternal#1#2{%
  \clap{$\mathsurround=0pt#1{#2}$}}

\let\oldstackrel\stackrel
\renewcommand{\stackrel}[2]{\oldstackrel{\mathclap{#1}}{#2}}

\newcommand{\parastoo}{\textcolor{black}}

\newcommand{\Roy}{\textcolor{red}}

\newcommand{\Royy}{\textcolor{black}}

\newcommand{\Roycr}{\textcolor{black}}

\newcommand{\Rarxiv}{\textcolor{black}}

\newcommand{\leak}{L}

\allowdisplaybreaks
\begin{document}

\title{Information Leakage in Index Coding}

\author{Yucheng Liu$^{\dag}$\thanks{This work was supported by the ARC Discovery Scheme DP190100770; the US National Science Foundation Grant CNS-1815322; and the ARC Future Fellowship FT190100429.}, Lawrence Ong$^{\dag}$, Phee Lep Yeoh$^{\S}$, Parastoo Sadeghi$^{\ddag}$, Joerg Kliewer$^{*}$, and Sarah Johnson$^{\dag}$, \\\vspace{-0mm}
$^{\dag}$The University of Newcastle, Australia (emails: \{yucheng.liu, \hspace{-0.5mm}lawrence.ong, \hspace{-0.5mm}sarah.johnson\}@newcastle.edu.au)\\
$^{\S}$University of Sydney, Australia (email: phee.yeoh@sydney.edu.au)\\
$^{\ddag}$University of New South Wales, Canberra, Australia (email: p.sadeghi@unsw.edu.au)\\
$^{*}$New Jersey Institute of Technology, USA (email: jkliewer@njit.edu)\\\vspace{0mm}
}

%\author{Yucheng Liu$^{\dag}$\thanks{This work was supported by the ARC Discovery Scheme DP190100770; the US National Science Foundation Grant CNS-1815322; and the ARC Future Fellowship FT190100429.}, Lawrence Ong$^{\dag}$, Sarah Johnson$^{\dag}$, Joerg Kliewer$^{*}$, Parastoo Sadeghi$^{\ddag}$, and Phee Lep Yeoh$^{\S}$\\\vspace{-0mm}
%$^{\dag}$The University of Newcastle, Australia (emails: \{yucheng.liu, \hspace{-0.5mm}lawrence.ong, \hspace{-0.5mm}sarah.johnson\}@newcastle.edu.au)\\
%$^{*}$New Jersey Institute of Technology, USA (email: jkliewer@njit.edu)\\
%$^{\ddag}$University of New South Wales, Australia (email: p.sadeghi@unsw.edu.au)\\
%$^{\S}$University of Sydney, Australia (email: phee.yeoh@sydney.edu.au)
%}

%\footnote{The work of P. Sadeghi was supported by the ARC Future Fellowship, FT190100429.}

%Emails:  $^{\dag}$\{yucheng.liu, parastoo.sadeghi\}@anu.edu.au, $^{*}$\{ni.ding, thierry.rakotoarivelo\}@data61.csiro.au}

%\author{Parastoo Sadeghi$^{\dag}$, Fatemeh Arbabjolfaei$^{*}$, and Young-Han Kim$^{*}$\\\vspace{-0mm}
%$^{\dag}$Research School of Engineering, Australian National University, Canberra, ACT, 2601, Australia\\
%$^{*}$Department of Electrical and Computer Engineering, University of California, San Diego, CA 92093, USA\\
%Emails:  parastoo.sadeghi@anu.edu.au, \{farbabjo, yhk\}@ucsd.edu}

\maketitle

%\footnote{The work of P. Sadeghi was supported by the ARC Future Fellowship, FT190100429.}

\begin{abstract}

We study the information leakage to a guessing adversary in index coding with a general message distribution. Under both vanishing-error and zero-error decoding assumptions, we develop lower and upper bounds on the optimal leakage rate, which are based on the broadcast rate of the subproblem induced by the set of messages the adversary tries to guess. When the messages are independent and uniformly distributed, the lower and upper bounds match, establishing an equivalence between the two rates.

\iffalse
We study the information leakage to a guessing adversary in zero-error source coding. The source coding problem is defined by a confusion graph capturing the distinguishability between source symbols. The information leakage is measured by the ratio of the adversary's successful guessing probability after and before eavesdropping the codeword, maximized over all possible source distributions. Such measurement under the basic adversarial model where the adversary makes a single guess and allows no distortion between its estimator and the true sequence is known as the maximum min-entropy leakage or the maximal leakage in the literature. 
We develop a single-letter characterization of the optimal normalized leakage under the basic adversarial model, together with an optimum-achieving scalar stochastic mapping scheme. An interesting observation is that the optimal normalized leakage is equal to the optimal compression rate with fixed-length source codes, both of which can be simultaneously achieved by some deterministic coding schemes. We then extend the leakage measurement to generalized adversarial models where the adversary makes multiple guesses and allows certain level of distortion, for which we derive single-letter lower and upper bounds. 
\fi

\end{abstract}

\section{Introduction}\label{sec:intro}

Index coding \cite{Birk--Kol1998,bar2011index} %, introduced by Birk and Kol in the context of satellite communication \cite{Birk--Kol1998}, 
studies the communication problem where a server broadcasts messages via a noiseless channel to multiple receivers with side information. 
Due to its simple yet fundamental model, 
%as well as its intrin
index coding has been recognized as a canonical problem in network information theory, and is closely connected with many other problems such as network coding, distributed storage, and coded caching. 
%Each receiver requests one unique message and has prior knowledge of some other messages as its side information. 
Despite substantial progress achieved so far (see \cite{arbabjolfaei2018fundamentals} and the references therein), the index coding problem remains open in general.

\iffalse
The index coding problem with security constraints was first studied by Dau et al. in \cite{dau2012security}, where in addition to the \emph{legitimate} receivers there is an eavesdropper who knows some messages as its side information and wants to obtain any other message. 
The server must broadcast in such a way that the legitimate receivers can decode their requested messages while the eavesdropper cannot decode any single message aside from the messages it already knows. 
Several extensions have been studied in \cite{ong2016secure,ong2018secure,mojahedian2017perfectly,liu:vellambi:kim:sadeghi:itw18}. 
%Such secure variant of the index coding problem in the presence of an eavesdropper is commonly referred to as \emph{secure index coding} in the literature. 
\fi

%Various notions of information leakage have been studied from graph-theoretic perspectives in the literature \cite{xxx}. 
%Recently, we have investigated the information leakage to a guessing adversary in zero-error source coding \cite{ddd}, where the source coding problem is defined by a confusion graph \cite{korner1973coding}. 
%In contrast, in this work, we study the information leakage in index coding, which is intrinsically different from \cite{ddd} in the following aspects. 

\begin{figure}[ht]
\begin{center}
\includegraphics[scale=0.1625]{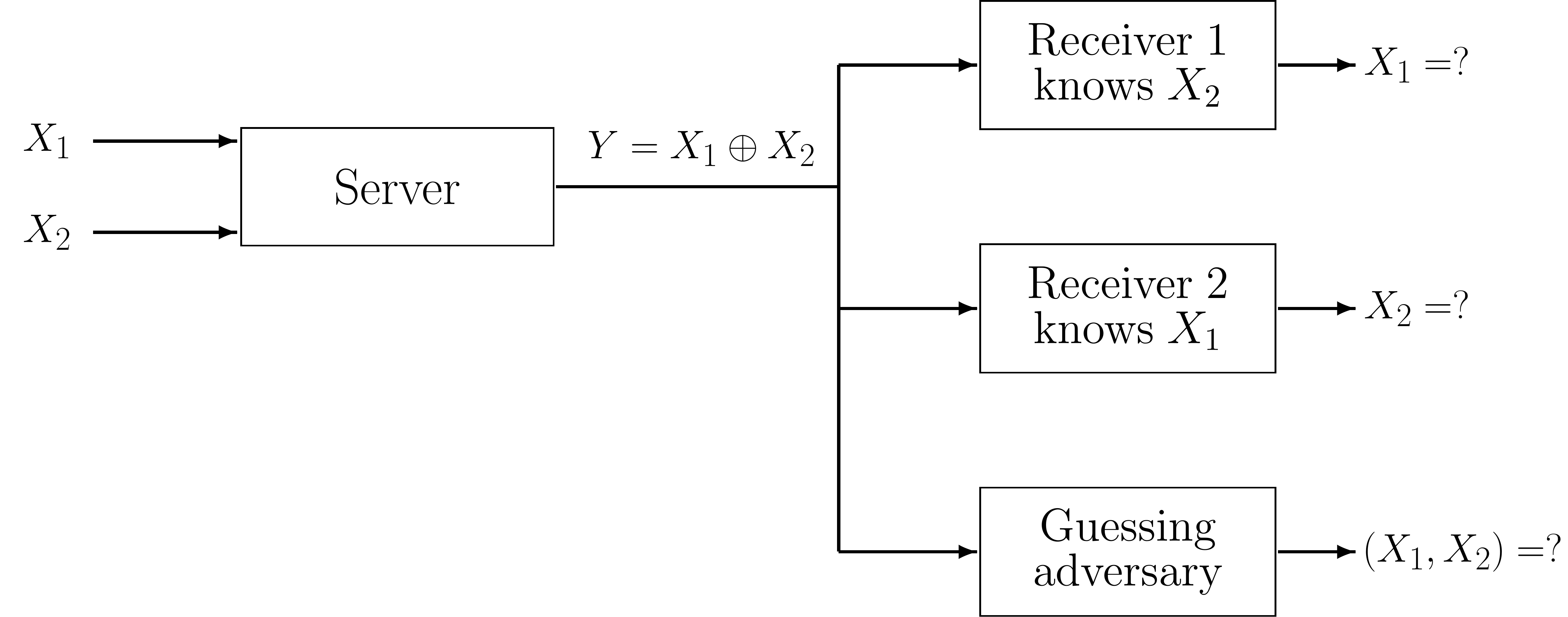}
\caption{There are two correlated binary messages $X_{\{1,2\}}=(X_1,X_2)$ with distribution: $P_{X_{\{1,2\}}}(0,0)=0.1$, $P_{X_{\{1,2\}}}(0,1)=0.2$, $P_{X_{\{1,2\}}}(1,0)=0.3$, and $P_{X_{\{1,2\}}}(1,1)=0.4$. 
A binary codeword $Y$ is generated by the server as $Y=X_1\oplus X_2$ and broadcast to the receivers. 
Every receiver can decode its wanted message based on the codeword and its side information. 
%according to the following mapping: $P_{Y|X_{1,2}}(y_1|0,0)=$. 
An adversary eavesdrops the codeword $Y$ and makes a single guess on $X_{\{1,2\}}$. 
If $Y=0$, the adversary's \Royy{guess} will be $(1,1)$ as $P_{X_{\{1,2\}}|Y}(1,1|0)=0.8>P_{X_{\{1,2\}}|Y}(0,0|0)=0.2>P_{X_{\{1,2\}}|Y}(0,1|0)=P_{X_{\{1,2\}}|Y}(1,0|0)=0$. 
Similarly, if $Y=1$, the adversary's \Royy{guess} will be $(1,0)$. %, which is the most likely realization given $Y$ being $1$. 
%as $P_{Y|X_{1,2}}(1|(1,0))=0.6>P_{Y|X_{1,2}}(0|(0,1))=0.4$. 
%according to the maximum likelihood estimation rule, the 
%Upon receiving codeword $y_1$, the adversary will guess $x_4=11$, upon receiving $y_2$, the adversary will guess $x_{1,2}=10$. 
%\Royy{add a toy example to show the behavior of the adversary.}
\vspace{-6mm}}
\label{fig:index:coding:with:adversary}
\end{center}
\end{figure}

%The security aspect of index coding has been studied in \cite{dau2012security,ong2016secure,liu:vellambi:kim:sadeghi:itw18,ong2021code}, where 
In secure index coding \cite{dau2012security,ong2016secure,liu:vellambi:kim:sadeghi:itw18,ong2021code}, 
%in addition to the legitimate receivers there is an adversary (eavesdropper) and 
the server must simultaneously satisfy the legitimate receivers' decoding requirements and protect the content of some messages from being obtained by an eavesdropping adversary. 
%An alternative setup where there is no adversary and the security constraints are against the receivers themselves was introduced in \cite{dau2012security} and later studied in \cite{liu2020secure,private:index:coding:arxiv}. 
A variant of this setup puts security constraints on the receivers themselves against some messages \cite{dau2012security,liu2020secure,private:index:coding:arxiv}.
%The authors of \cite{karmoose2019privacy} considered the index coding problem from a different privacy-preserving perspective, where the goal is to limit the information that a receiver can infer about the \emph{identities} of the requests of other receivers. 
Instead of protecting the messages, another variant of index coding has also been studied from a privacy-preserving perspective, where the goal is to limit the information that a receiver can infer about the \emph{identities} of the requests of other receivers \cite{karmoose2019privacy}. 
The privacy-utility tradeoff in a multi-terminal data publishing problem inspired by index coding was investigated in \cite{yuchengliu2020isit}. 
In this work, we study the information leakage to a guessing adversary in index coding, which, to the best of our knowledge, has not been considered in the literature. 
The adversary eavesdrops the broadcast codeword and tries to guess the message tuple via maximum likelihood estimation within a certain number of trials. 
%The leakage to the adversary is quantified as the ratio between the adversary’s probability of successful guessing \emph{after and before} observing the codeword. 
Our aim is to characterize the information leakage to the adversary, which is defined as the ratio between the adversary’s probability of successful guessing \emph{after and before} observing the codeword \cite{smith2009foundations,braun2009quantitative,issa2019operational}. 
%Such way of measuring information leakage has been introduced and investigated in \cite{smith2009foundations,braun2009quantitative,issa2018operational}. 
For a visualization of the problem setup, see Figure \ref{fig:index:coding:with:adversary}.

Recently we have studied \cite{isit:2021:arxiv} information leakage to a guessing adversary in zero-error source coding defined by a family of confusion graphs \cite{korner1973coding}. 
While the index coding problem can also be characterized by a confusion graph family \cite{alon2008broadcasting}, the study of information leakage in index coding is intrinsically different from that of source coding in the following aspects. 
The most significant difference comes from the different internal structures within the two confusion graph families. 
More specifically, for the source coding model we considered \cite{isit:2021:arxiv}, the relationship among the confusion graphs of different sequence lengths is characterized by the {disjunctive product} \cite{scheinerman2011fractional}. 
On the other hand, for the index coding problem, the relationship among the confusion graphs cannot be characterized by any previously defined graph product. 
Another difference is that while our previous work \cite{isit:2021:arxiv} requires zero-error decoding at the legitimate receiver assuming \emph{worst-case} source distribution, this paper considers both zero-error and vanishing-error scenarios and assume a \emph{general} message distribution. 
Furthermore, in this work we take into account the adversary's side information which can include any message in the system.

Our main contribution is developing lower and upper bounds (i.e., converse and achievability results) on the optimal information leakage rate, for both vanishing-error and zero-error scenarios. 
The converse bound is derived using graph-theoretic techniques based on the notion of confusion graphs for index coding \cite{alon2008broadcasting}. 
The achievability result is established by constructing a deterministic coding scheme as a composite of the coding schemes for two subproblems, one induced by the messages the adversary knows as side information and the other induced by the messages the adversary does not know and thus tries to guess. 

Moreover, we show that when the messages are uniformly distributed and independent of each other (as in most existing works for index coding), 
%or when the server only cares about BLABLA, 
the lower and upper bounds developed match. This establishes an equivalence between the optimal leakage rate of the problem and the optimal compression rate of the subproblem induced by the messages the adversary tries to guess. 

\section{System Model and Problem Formulation}\label{sec:model}

%\iffalse

%\Royy{This section is copied from ISIT 2021 paper; to be changed}

%%{\textbf{Source coding with confusion graph $\Gamma$:}

{\it Notation:} For any $a\in {\mathbb Z}^+$, $[a]\doteq \{1,2,\cdots,a\}$. 
%For non-negative integers $a$ and $b$, $[a]\doteq \{1,2,\cdots,a\}$, and $[a:b]\doteq \{a,a+1,\cdots,b\}$. If $a>b$, $[a:b]=\emptyset$. 
%For a set $S$, $|S|$ denotes its cardinality and $2^S$ denotes its power set. 
%{\it Notation:} For non-negative integers $a$ and $b$, $[a]$ denotes the set $\{1,2,\cdots,a\}$, and $[a:b]$ denotes the set $\{a,a+1,\cdots,b\}$. If $a>b$, $[a:b]=\emptyset$. For a set $S$, $|S|$ denotes its cardinality and $2^S$ denotes its power set. 
%For any given index coding problem, we use $n$ to denote the number of messages in the problem, and $N\doteq 2^{[n]}$ to denote the 
For any discrete random variable $Z$ with probability distribution $P_Z$, we denote its alphabet by $\Zc$ with realizations $z\in \Zc$. 
%For any $K\subseteq \Zc$, $P_Z(K)\doteq \sum_{z\in K}P_Z(z)$. 

There are $n$ discrete memoryless stationary messages (sources), $X_i,i\in [n]$, of some common finite alphabet $\Xcal$. 
%Assume there are $n$ messages $X_i,i\in [n]$ of some common discrete finite alphabet $\Xcal$. 
For any $S\subseteq [n]$, set $\Xv_S\doteq (X_i,i\in S)$, $\xv_S\doteq (x_i,i\in S)$, and $\Xc_S\doteq \Xc^{|S|}$. 
Thus $X_{[n]}$ denotes the tuple of all $n$ messages, and $x_{[n]}\in \Xc_{[n]}$ denotes a realization of the message $n$-tuple. 
By convention, $X_{\emptyset} = x_{\emptyset} = \Xc_{\emptyset} = \emptyset$. 
We consider an arbitrary, but fixed distribution $P_{X_{[n]}}$ on $\Xc_{[n]}$, assuming without loss of generality that it has full support\footnote{\Royy{While a common assumption in most index coding literatures is that the messages are independent and uniformly distributed, here we consider the more general case with arbitrary joint distribution for the messages.}}.

There is a server containing all messages. 
It encodes the tuple of $n$ message sequences $X^t_{[n]}=(X^t_i,i\in [n])$ according to some (possibly randomized) encoding function $f$ to some codeword $Y$ that takes values in the code alphabet $\Yc=\{1,2,\ldots,M\}$. 
Each message sequence $X^t_i=(X_{i,1},X_{i,2},\ldots,X_{i,t})$ is of length $t$ symbols.  
%That is, $Y=f(X^t_{[n]})$. 
The server then transmits the codeword to $n$ receivers via a noiseless broadcast channel of normalized unit capacity. 
Let $P_{Y,X^t_{[n]}}$ denote the joint distribution of the message sequence tuple $X^t_{[n]}$ and the codeword $Y$. 
For any $S\subseteq [n]$, we define the following notation for message sequence tuples.
\begin{itemize}
\item $\Xc^t_S=\Xc^{t|S|}$;
\item $X^t_S=(X^t_i,i\in S)=(X_{S,1},X_{S,2},\ldots,X_{S,t})$, where $X_{S,j}=(X_{i,j},i\in S)$ for every $j\in [t]$. Note that $X_{i,j}$ denotes the $j$-th symbol of message sequence $X^t_i$. 
\item Similarly, $x^t_S=(x^t_i,i\in S)=(x_{S,j},j\in [t])$, where $x_{S,j}=(x_{i,j},i\in S)$ for every $j\in [t]$. 
\end{itemize}
%and $x^t_S$ and $\Xc^t_S$ can be defined similarly. 
Also, as the messages are memoryless, for any $x^t_{[n]}=(x_{[n],1},x_{[n],2},\ldots,x_{[n],t})$, $P_{X^t_{[n]}}(x^t_{[n]})=\prod_{j\in [t]}P_{X_{[n]}}(x_{[n],j})$. 

On the receiver side, 
%each receiver wants to learn one unique message and knows some other messages already as side information. 
%More specifically, 
we assume that receiver $i \in [n]$ wishes to obtain message $X^t_i$ and knows $X^t_{A_i}$ as side information for some $A_i \subseteq [n]\setminus \{i\}$. 
%receiver $i \in [n]$  knows $X^t_{A_i}$ as side information for some $A_i \subseteq [n]\setminus \{i\}$, and 

More formally, a $(t,M,f,\gv)$ {\em index code} can be defined by
%we define a $(t,M,f,\gv)$ {\em index code} by
\begin{itemize}
\item One stochastic encoder $f: \Xc^{nt} \to \{1,2,\ldots,M\}$ at the server that maps each message sequence tuple $\xv^t_{[n]}\in \Xc^{nt}$ to a codeword $y\in \{1,2,\ldots,M\}$, and
\item $n$ deterministic decoders $\gv=(g_i,i\in [n])$, one for each receiver $i \in [n]$, such that $g_i: \{1,2,\ldots,M\} \times \Xc^{t|A_i|} \to \Xc^{t}$ maps the codeword $y$ and the side information $\xv^t_{A_i}$ to some estimated sequence $\hat{x}^t_i$.
\end{itemize}

For any $\e>0$, we say a $(t,M,f,\gv)$ index code is \emph{valid} (with respect to $\e$) if and only if (iff) the average probability of error satisfies ${\rm P}_{\rm e}\doteq {\rm P}\{ (\Xh_{[n]}) \ne (X_{[n]})\} \le \epsilon$. 
%\begin{equation}
%{\rm P}_{\rm e}\doteq {\rm P}\{ (\Xh_1,\ldots, \Xh_n) \ne (X_1, \ldots, X_n)\} \le \epsilon.  \label{eq:model:vanishing:error}
%\end{equation}
%
%
We say a compression rate $R$ is achievable iff for every $\epsilon > 0$, there exists a valid $(t,M,f,\gv)$ code such that $R\ge (\log M)/t$. 
%\begin{equation}
%R\ge \log M/t.  \label{eq:model:communication:rate}
%\end{equation}

The \emph{optimal} compression rate $\Rc$, also referred to as the \emph{broadcast rate}, can be defined as 
\begin{align}  \label{eq:model:broadcast:rate}
\Rc%&\doteq \lim_{\e\to 0}\inf_{t}\inf_{\substack{\text{valid $(t,M,f,\gv)$ code}}}\frac{\log M}{t}  \nonumber  \\
&=\lim_{\e\to 0}\lim_{t\to \infty}\inf_{\substack{\text{valid $(t,M,f,\gv)$ code}}}\frac{\log M}{t}. 
\end{align}
%where (a) can be shown using Fekete's lemma \cite{fekete1923lemma}. 

%\Royy{We use $\rho$ to denote the zero-error broadcast rate.}

We say a $(t,M,f,\gv)$ \Royy{code} is valid with respect to zero-error decoding iff the average probability of error is zero. 
%The zero-error optimal communication rate $\rho$, also referred to as the zero-error broadcast rate, is defined as 
The zero-error broadcast rate $\rho$ can then be defined as 
\begin{align}  \label{eq:model:zero:error:broadcast:rate}
\rho\doteq %\inf_{t}\inf_{\substack{\text{$(t,M,f,\gv)$ code}}}\frac{\log M}{t}=
\lim_{t\to \infty}\inf_{\substack{\text{valid $(t,M,f,\gv)$ code w.r.t.}\\ \text{zero-error decoding}}}\frac{\log M}{t}.
\end{align}
%\begin{equation}
%{\rm P}_{\rm e}={\rm P}\{ (\Xh_1,\ldots, \Xh_n) \ne (X_1, \ldots, X_n)\}=0.  \label{eq:model:zero:error}
%\end{equation}

Clearly, by definition, we always have $\Rc\le \rho$.

%Any index coding problem can be represented by a sequence $(i|A_i),i\in [n]$, specifying the side information availability at receivers. 
The side information availability at receivers for a specific index coding instance can be represented by a sequence $(i|j\in A_i),i\in [n]$. %\cite{arbabjolfaei2018fundamentals}
%For example, the 
%Alternatively, any index coding instance can be characterized by 
Alternatively, it can be characterized by 
a family of \emph{confusion graphs}, $(\Gamma_t,t\in {\mathbb Z}^+)$ \cite{alon2008broadcasting}. 
For a given sequence length $t$, the confusion graph $\Gamma_t$ is an undirected graph defined on the message sequence tuple alphabet $\Xc^t_{[n]}$. That is, $V(\Gamma_t)=\Xc^t_{[n]}$. 
%When we say vertex $x^t_{[n]}$ in $\Gamma$, we mean the vertex corresponding to the realization $x^t_{[n]}$. 
Vertex $x^t_{[n]}$ in $\Gamma$ corresponds to the realization $x^t_{[n]}$. 
Any two different vertices $x^t_{[n]},z^t_{[n]}$ are adjacent in $\Gamma_t$ iff $x^t_i\neq z^t_i$ and $x^t_{A_i}=z^t_{A_i}$ for some receiver $i\in [n]$. We call any pair of vertices satisfying this condition \emph{confusable}. 
Hence, $E(\Gamma_t)=\{ \{x^t_{[n]},z^t_{[n]}\}:\text{$x^t_i\neq z^t_i$ and $x^t_{A_i}=z^t_{A_i}$ for some $i\in [n]$} \}$. 
%their corresponding message tuple realizations are confusable. 

\begin{figure}[ht]
\begin{center}
\includegraphics[scale=0.35]{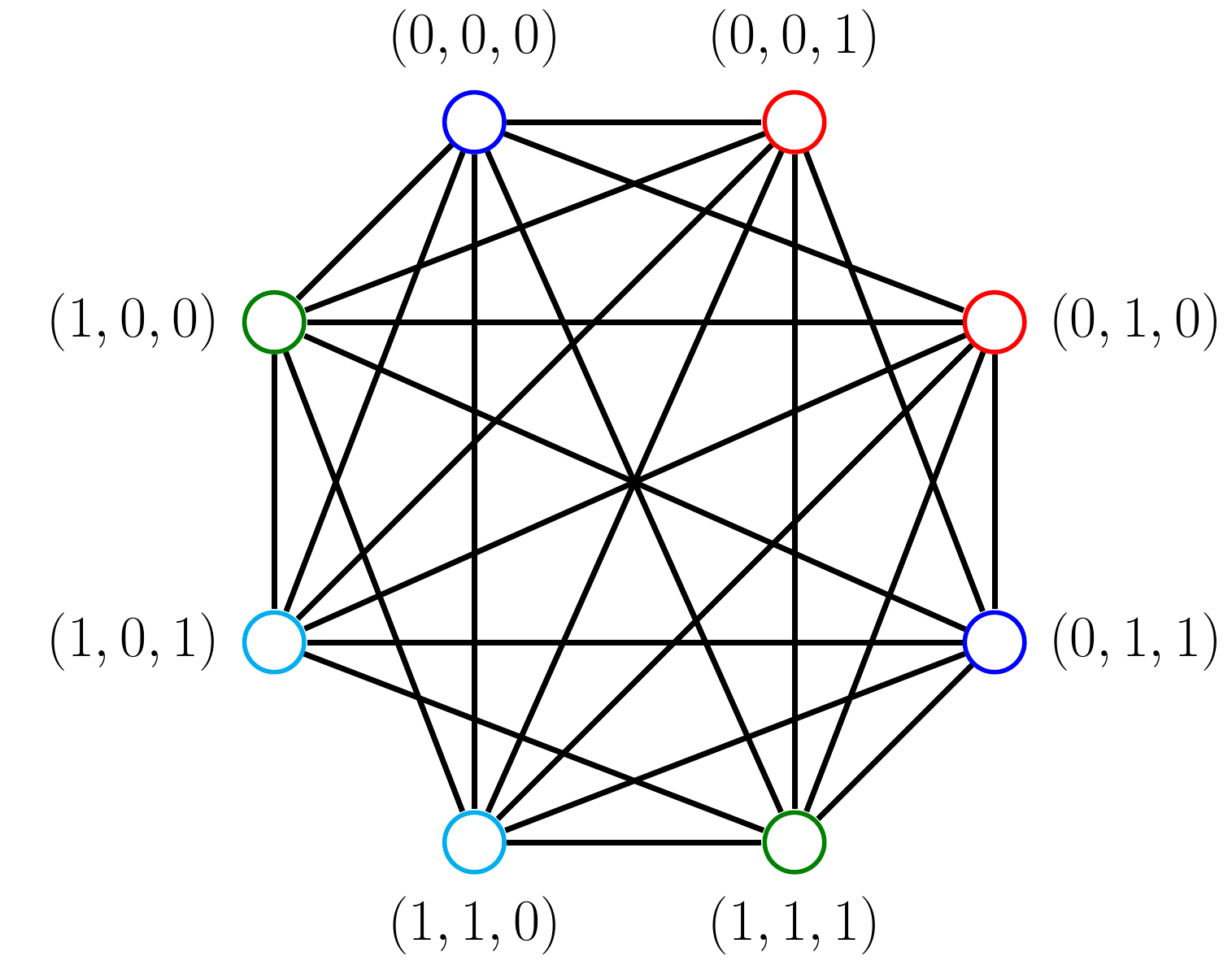}
\caption{The confusion graph $\Gamma_1$ with $t=1$ for the $3$-message index coding instance $(1|-),(2|3),(3|2)$. Note that, for example, $x_{[n]}=(0,0,0)$ and $z_{[n]}=(0,0,1)$ are confusable because $x_3=0\neq z_3=1$ and $x_{A_3}=x_2=0=z_2=z_{A_3}$. Suppose $(0,0,0)$ and $(0,0,1)$ are mapped to the same codeword $y$ with certain nonzero probabilities. Then upon receiving this $y$, receiver $3$ will not be able to tell whether the value for $X_3$ is $0$ or $1$ based on its side information of $X_2=0$. 
For this graph, it can be easily verified that the independence number is $2$, and that the chromatic number equals to the fractional chromatic number, both of which equal to $4$. 
%For this graph, it can be easily verified that the independence number $\alpha(\Gamma_1)$ is $2$, and that the chromatic number $\chi(\Gamma_1)$ equals to the fractional chromatic number $\chi_{\rm f}(\Gamma_1)$, both of which equal to $4$. 
We have drawn an optimal coloring scheme with $4$ colors in the graph. 
%For basic 
}
\label{fig:confusion:graph}
\end{center}
\vspace{-4mm}
\end{figure}

For correct decoding at all receivers, any two realizations $x^t_{[n]},z^t_{[n]}$ can be mapped to the same codeword $y$ with nonzero probabilities iff they are not confusable \cite{alon2008broadcasting}. See Figure \ref{fig:confusion:graph} below for a toy example of an index coding instance and its confusion graph. 
For the definitions for basic graph-theoretic notions, see any textbook on graph theory (e.g., Scheinerman and Ullman \cite{scheinerman2011fractional}). 
%see \cite{scheinerman2011fractional}.

Consider any set $S\subseteq [n]$. 
The subproblem induced by $S$ is jointly characterized by the distribution $P_{X_S}$ and the sequence $(i|A_i\cap S),i\in S$. %\footnote{More rigorously, one needs to define an arbitrary bijective function $\eta:[|S|]\to S$ and the subproblem induced by $S$ will be jointly characterized by $P_{X_S}$ and $(i|\{ j\in [S]:\eta(j)\in S\cap A_{\eta(i)} \}),i\in [|S|]$.}
Let $\Gamma_t(S)$ denote the confusion graph of sequence length $t$ of the subproblem induced by $S$. 
Let $\Rc(S)$ and $\rho(S)$ denote the broadcast rate and zero-error broadcast rate of the subproblem induced by $S$, respectively. 

%\begin{example}
%\Royy{add a toy example if we have enough space}
%\end{example}

{\textbf{Preliminaries on $\Rc$ and $\rho$:}}
Consider any index coding problem characterized by confusion graphs $(\Gamma_t,t\in {\mathbb Z}^+)$ and distribution $P_{X_{[n]}}$. %$(i|A_i),i\in [n]$. 
%We start with the following simple lemma, the proof of which is relegated to Appendix \ref{app:proof:lem:deterministic:enough:for:compression}. 
We start with the following lemma. %, which justifies the assumption of deterministic encoding functions in existing literature
\begin{lemma}  \label{lem:deterministic:enough:for:compression}
To characterize the broadcast rate $\Rc$ and zero-error broadcast rate $\rho$, it suffices to only consider index codes with deterministic encoding function $f$. 
\end{lemma}

The above lemma can be simply proved by showing that given any valid index code with a stochastic encoding function, one can construct another valid code with a deterministic encoding function and same or smaller compression rate. %We omit the proof details due to limited space. 

Most existing results in the literature on the optimal compression rate (vanishing or zero error) of index coding were established assuming deterministic encoding functions. 
Lemma \ref{lem:deterministic:enough:for:compression} indicates that those results can be directly applied to characterizing $\Rc$ and $\rho$. 
%\fi
%The above lemma thus justifies the assumption of deterministic encoding functions in existing literature. 

Since we are considering fixed-length codes (rather than variable-length codes), the zero-error broadcast rate $\rho$ does not depend on $P_{X_{[n]}}$ and can be characterized solely by the confusion graphs $(\Gamma_t,t\in {\mathbb Z}^+)$ \cite{alon2008broadcasting} as
\begin{align}
\rho=\lim_{t\to \infty}\frac{1}{t}\log \chi(\Gamma_t)\stackrel{(a)}{=}\lim_{t\to \infty}\frac{1}{t}\log \chi_{\rm f}(\Gamma_t),  \label{eq:compression:zero:error}
\end{align}
where $\chi(\cdot)$ and $\chi_{\rm f}(\cdot)$ respectively denote the chromatic number and fractional chromatic number of a graph, %(see \cite{scheinerman2011fractional} for more details)
and the proof of $(a)$ can be found in \cite[Section 3.2]{arbabjolfaei2018fundamentals}. 
%See \cite[Section 2]{arbabjolfaei2018fundamentals} for more details on basic graph-theoretic definitions. 

It has been shown \cite{Langberg--Effros2011} that, with the messages $X_{[n]}$ being uniformly distributed and independent of each other, the vanishing-error broadcast rate $\Rc$ equals to the zero-error broadcast rate $\rho$. 
%That is, 
%\begin{align}
%\Rc=\rho, \qquad \text{If $P_{X_{[n]}}$ is uniform distribution.}
%\end{align}
Such equivalence does not hold for a general distribution $P_{X_{[n]}}$ as it has been shown in \cite{miyake2015index} that the (vanishing-error) broadcast rate $\Rc$ can be strictly smaller than its zero-error counterpart $\rho$.

\iffalse
We briefly review some known results on compression rate $\beta^*$ and $\lambda^*$ \Royy{we use $\lambda^*$ to denote the broadcast rate under zero-error decoding}. 
It has been shown in \cite{miyake2015index} that the vanishing-error compression rate $\beta^*(\Gamma,P_{X_{[n]}})$ can be strictly smaller than its zero-error counterpart $\lambda^*(\Gamma,P_{X_{[n]}})$. 
The latter is equal to $\lambda^*(\Gamma,U_{X_{[n]}})$, i.e., the broadcast rate under uniform message tuple distribution, since zero-error compression rate with fixed-length index codes is independent of the underlying distribution. 

With uniformly distributed independent sources, it has been shown in \cite{Langberg--Effros2011} that the zero-error broadcast rate is equal to the vanishing-error broadcast rate. 
%Thus in this subsection we may refer to both the zero-error and vanishing-error compression rate simply as the compression rate. 

The zero-error compression rate $\lambda^*(\Gamma)$ and vanishing-error compression rate $\beta^*(\Gamma)$ are equal and can be expressed based on the confusion graph $\Gamma$ as 
\begin{align}
\lambda^*(\Gamma)=\beta^*(\Gamma)=\lim_{t\to \infty}\frac{1}{t}\log \chi(\Gamma_t)\stackrel{(a)}{=}\lim_{t\to \infty}\frac{1}{t}\log \chi(\Gamma_t),  \label{eq:index:coding:compression:uniform}
\end{align}
where $(a)$ has been shown in \cite[Section 3.2]{arbabjolfaei2018fundamentals}.
\fi

%\Royy{The following needs to be revised to include guessing capability function $g(t)$ and adversary's side information $A_e$}

{\textbf{Leakage to a guessing adversary:}}
%As a starting point, we assume that the adversary makes a single guess after observing each codeword and allows no distortion between its estimator sequence and the true source sequence. 

We assume the adversary knows messages $X_{P}$ and tries to guess the remaining messages $X_{Q}$, where $Q=[n]\setminus P$, via maximum likelihood estimation within a number of trials. 
\Royy{In other words, the adversary generates a list of certain size of guesses, and is satisfied iff the true message sequence is in the list}. 
%In other words, the adversary always guesses the most probable message tuple realization, and if not correct, then the second most probable realization, and so on, until it exhausts the number of guesses it can make. 
%\Royy{For example, if the adversary possesses certain testing mechanism to check whether each estimation is correct, then it will always guess the most probable realization, and if not correct, then the second most probable realization, and so on, until it exhausts the number of guesses it can make}. 
We characterize the number of guesses the adversary can make by a function of sequence length, $c:{\mathbb Z}^+\to {\mathbb Z}^+$, namely, the guessing capability function. 
%a \emph{guessing capability} function $c(t)$, where $t\in {\mathbb Z}^+$ is the sequence length. 
We assume $c(t)$ to be non-decreasing and upper-bounded\footnote{It can be verified that if for some $t$ we have $c(t)>\alpha(\Gamma_t(Q))$, then the probability of the adversary successfully guessing $x^t_Q$ after observing $y$ is at least $1-{\rm P}_{\rm e}$, which tends to $1$ as $\epsilon$ tends to $0$, making the problem trivial. 
%then the adversary can successfully guess $x^t_Q$ with probability arbitrarily close to $1$.
}
%\footnote{Suppose for some $t$ we have $c(t)\ge \alpha(\Gamma_t)$. Then upon observing any codeword $y$, the adversary can always determine the true source value by exhaustively guessing all possible $x^t\in \Xc^t(y)$ as $|\Xc^t(y)|\le \alpha(\Gamma_t)$.}
by $\alpha(\Gamma_t(Q))$, 
where $\alpha(\cdot)$ denotes the independence number of a graph. %\cite{scheinerman2011fractional}. 

Consider any valid $(t,M,f,\gv)$ index code. 
Before eavesdropping the codeword $y$, the expected probability of the adversary successfully guessing $x^t_Q$ with $c(t)$ number of guesses is 
\begin{align*}
{\rm P_{\rm s}}(X^t_P)={\mathbb E}_{X^t_P} \left[ \max_{K\subseteq \Xc^t_Q:|K|\le c(t)} \sum_{x^t_Q\in K} P_{X^t_Q|X^t_P}(x^t_Q|X^t_P) \right],
\end{align*}
and the expected successful guessing probability after observing $y$ is 
\begin{align*}
{\rm P_{\rm s}}(X^t_P,Y) \hspace{-0.25mm} = \hspace{-0.25mm} {\mathbb E}_{Y,X^t_P} \hspace{-0.5mm} \left[ \max_{\substack{K\subseteq \Xc^t_Q:\\|K|\le c(t)}} \sum_{x^t_Q\in K} \hspace{-0.5mm} P_{X^t_Q|Y,X^t_P}(x^t_Q|Y,X^t_P) \right] \hspace{-0.5mm} .
\end{align*}
The leakage to the adversary, denoted by $\leak$, is defined as the logarithm of the ratio between the expected probabilities of the adversary successfully guessing $x_Q$ \emph{after and before} observing the transmitted codeword $y$. That is, 
\begin{align}
\leak\doteq \log \frac{{\rm P_{\rm s}}(X^t_P,Y)}{{\rm P_{\rm s}}(X^t_P)}.  \label{eq:def:leakage:rate:any:t:f}
\end{align}
%\begin{align}
%&L(f)  \nonumber  \\
%&\doteq \log \frac{{\rm P_{\rm s}}(X_P,Y)}{{\rm P_{\rm s}}(X_P)}  \nonumber  \\
%%&\doteq \log \frac{ {\mathbb E}_{Y,X_P} \left[ \max\limits_{K\subseteq \Xc_Q:|K|\le c(t)} \sum\limits_{x_Q\in K} P_{X_Q|Y,X_P}(x_Q|Y,X_P) \right] }{ {\mathbb E}_{X_P} \left[ \max\limits_{K\subseteq \Xc_Q:|K|\le c(t)} \sum\limits_{x_Q\in K} P_{X_Q|X_P}(x_Q|X_P) \right] }  \nonumber  \\%\label{eq:def:maxl:leakage:rate:any:t:mapping:operational}  \\
%%%%
%%&=\log \max_{P_X} \frac{\sum\limits_{y\in \Yc} \max\limits_{x^t\in \Xc^t} P_{X^t,Y}(x^t,y)}{\max\limits_{x^t\in \Xc^t}P_{X^t}(x^t)}  \label{eq:def:max:leakage:rate:any:t:mapping}  \\
%%&=\log \sum\limits_{y\in \Yc} \max\limits_{x^t\in \Xc^t} P_{Y|X^t}(y|x^t),\label{eq:def:max:leakage:rate:simplified:any:t:mapping}
%%%%
%&=\log \frac{ \sum\limits_{x_P\in \Xc_P,y\in \Yc} \max\limits_{K\subseteq \Xc_Q:|K|\le c(t)} \sum\limits_{x_Q\in K} P_{X_{[n]},Y}(x_P,x_Q,y) }{ \sum\limits_{x_P\in \Xc_P} \max\limits_{K\subseteq \Xc_Q:|K|\le c(t)} \sum\limits_{x_Q\in K} P_{X_{[n]}}(x_P,x_Q) },  \label{eq:def:leakage:any:f}
%\end{align}
%where $\eqref{eq:def:max:leakage:rate:simplified:any:t:mapping}$ follows from \cite[Proposition 5.1]{braun2009quantitative}. %, which is equal to the Sibson mutual information of order infinity \cite{verdu2015alpha} betwen $X^t$ and $Y$. %, $I_{\infty}^{\rm Sibson}(X^t;Y)$ . 
%
%
The (optimal) leakage rate can then be defined as 
\begin{align}
\Lc&\doteq \lim_{\e\to 0}\lim_{t\to \infty} t^{-1} \inf_{\text{valid $(t,M,f,\gv)$ code}} \leak.  \label{eq:def:leakage:rate} 
\end{align}
%
%
\iffalse
The \emph{optimal} leakage for a given message length $t$ is 
\begin{align}
\Lc^t&\doteq \inf_{\text{all valid $(t,r,f,\gv)$ index codes}} L^t(f)\label{eq:def:leakage:rate:any:t},  
\end{align}
based upon which we define the (optimal) leakage rate as 
%and then the (optimal) leakage rate 
\begin{align}
\Lc&\doteq \lim_{t\to \infty} t^{-1} \Lc^t. \label{eq:def:leakage:rate} 
\end{align}
\fi

\begin{remark}
\Royy{It can be readily verified that the leakage metric $\leak$ is always non-negative. 
When $c(t)=1$ (i.e., the adversary only makes a single guess after each observation), $\leak$ reduces to the min-entropy leakage \cite{smith2009foundations}. When $c(t)=1$ and the messages are uniformly distributed, $\leak$ is equal to the maximal leakage \cite{issa2019operational} and the maximum min-entropy leakage \cite{braun2009quantitative}}. 
%When $c(t)=1$ (i.e., the adversary only makes a single guess after each observation) and the messages are uniformly distributed, the leakage metric $\leak$ is equal to the maximal leakage \cite{issa2019operational} from $X_{[n]}$ to $Y$, which is also equal to the maximum min-entropy leakage \cite{braun2009quantitative} from $X_{[n]}$ to $Y$. 
\end{remark}

If we require zero-error decoding at receivers, the zero-error (optimal) leakage rate $\lambda$ can be similarly defined as 
%\Royy{We use $\lambda$ to denote the zero-error leakage rate.}
\begin{align}
\lambda&\doteq \lim_{t\to \infty} t^{-1} \inf_{\substack{\text{valid $(t,M,f,\gv)$ code w.r.t.}\\ \text{zero-error decoding}}} \leak.  \label{eq:def:zero:error:leakage:rate} 
\end{align}

By definition, we always have $\Lc\le \lambda$.

\section{Information Leakage in Index Coding}  \label{sec:index:coding}

%Consider an index coding problem where a sender knows $n$ messages $X_1,X_2,\ldots,X_n$, which are uniformly distributed and independent of each other, and transmits a codeword $Y$, generated according to mapping $P_{Y|X_{[n]}}$, to $n$ receivers through a noiseless broadcast channel. Receiver $i$ wishes to decode message $X_i$ without error and knows messages $X_{A_i}$ as side information for some $A_i\subseteq [n]\setminus \{i\}^c$. 
%Such problem can be represented by its corresponding confusion graph $\Gamma$. Thus we may simply refer to an index coding problem by its confusion graph $\Gamma$. 
%For more details, see \cite[Section 3.1]{arbabjolfaei2018fundamentals}. 
%%a directed graph $G$ called side information graph, where 

%\Royy{this description to be revised}

%In this section, we consider three different scenarios regarding the probability distribution of the messages: 
%\begin{enumerate}
%\item The messages are uniformly distributed and independent of each other, which is the most common assumption used in most index coding literature. 
%\item The messages have a general joint distribution $P_{X_{[n]}}$, which is unknown to the sender. Consequently, the sender is interested in the worst-case compression rate and leakage rate. 
%\item The messages have a general joint distribution $P_{X_{[n]}}$ and it is known to the sender. 
%\end{enumerate}
%In each scenario, we will consider both the zero-error setting and the vanishing-error setting. 

\subsection{Leakage Under A General Message Distribution}

%In this subsection, we assume that the messages have a general joint distribution $P_{X_{[n]}}$ known to the sender. 

%Consider any index coding problem $\Gamma$ with message tuple distribution $P_{X_{[n]}}$. 
Consider any index coding problem $(i|j\in A_i),i\in [n]$ with confusion graphs $(\Gamma_t,t\in {\mathbb Z}^+)$ and distribution $P_{X_{[n]}}$. 
%In the following theorem, we propose lower and upper bounds for the leakage rate for both vanishing-error and zero-error cases.  
Our main result is the following theorem. 

\begin{theorem}  \label{thm:general}
For the vanishing-error leakage rate $\Lc$, we have
\begin{align}
\rho(Q)-|Q|+\log \frac{1}{\sum\limits_{x_{P}}\max\limits_{x_{Q}}P_{X_{[n]}}(x_{[n]})} \le \Lc\le \Rc(Q).  \label{eq:index:coding:Ae:multi:guess:general:vanishing}
\end{align}
%\begin{align}
%&\rho(Q)-|Q|+\log \frac{1}{\sum_{x_{P}}\max_{x_{Q}}P_{X_{[n]}}(x_{[n]})}  \nonumber  \\
%&\qquad \qquad \le \Lc\le \Rc(Q).  \label{eq:index:coding:Ae:multi:guess:general:vanishing}
%\end{align}
For the zero-error leakage rate $\lambda$, we have 
\begin{align}
&\rho(Q)-|Q|+\log \frac{1}{\sum\limits_{x_{P}}\max\limits_{x_{Q}}P_{X_{[n]}}(x_{[n]})} \le \lambda \le \rho(Q).  \label{eq:index:coding:Ae:multi:guess:general:zero:error}
\end{align}
%\begin{align}
%&\rho(Q)-|Q|+\log \frac{1}{\sum_{x_{P}}\max_{x_{Q}}P_{X_{[n]}}(x_{[n]})}   \nonumber  \\
%&\qquad \qquad\le \lambda \le \rho(Q).  \label{eq:index:coding:Ae:multi:guess:general:zero:error}
%\end{align}
\end{theorem}

%We first show the following lemma. 

In the following, we prove the lower and upper bounds in \eqref{eq:index:coding:Ae:multi:guess:general:vanishing}. 
As for \eqref{eq:index:coding:Ae:multi:guess:general:zero:error}, the lower bound follows directly from the lower bound in \eqref{eq:index:coding:Ae:multi:guess:general:vanishing} and the fact that $\Lc\le \lambda$, 
and the upper bound can be shown using similar techniques to the proof of the upper bound in \eqref{eq:index:coding:Ae:multi:guess:general:vanishing}. 

%\subsubsection{Converse result}

%We first show the lower bound in \eqref{eq:index:coding:Ae:multi:guess:general:vanishing}. 
\begin{IEEEproof}[Proof of the lower bound in \eqref{eq:index:coding:Ae:multi:guess:general:vanishing}]
%As we always have $\Lc_{g,A_e}^*\le \Lc_{g,A_e}$, it suffices to show that $\Lc_{g,A_e}^*\ge \beta(\Gamma(A_e^c))$ and that $\Lc_{g,A_e}\le \beta(\Gamma(A_e^c))$. 
%To show the former, 
%Consider any message length $t$, any decoding error $0<\epsilon<1$, and any mapping scheme $P_{Y|X^t_{[n]}}$ such that $P_e\le \epsilon$. 
Consider any $\e>0$ and any valid $(t,M,f,\gv)$ index code for which ${\rm P}_{\rm e}\le \e$. 

Consider any codeword $y\in \Yc$ and any realization %of the adversary's side information 
$x^t_P\in \Xc^t_P$. 
%Let $\Xc^t_Q(y,x^t_P,{\rm good})$ denotes the collection of $x^t_Q$ such that $(y,x^t_P,x^t_Q)$ are jointly possible and, with $x^t_{[n]}=(x^t_P,x^t_Q)$ being the true message sequence tuple realization and $y$ being the codeword, every receiver can correctly decode its requested message. %can be correctly decoded at every receiver.
Let $G_{\Xc^t_Q}(y,x^t_P)$ denote the collection of realizations $x^t_Q$ such that $(y,x^t_P,x^t_Q)$ \Royy{has nonzero probability}, and for the event that $x^t_{[n]}=(x^t_P,x^t_Q)$ is the true message sequence tuple realization and $y$ is the codeword realization, every receiver can correctly decode its requested message. %(i.e., $g_i(y,x^t_{A_i})=x^t_i, i\in [n]$). %can be correctly decoded at every receiver.
That is, 
\begin{align*}
G_{\Xc^t_Q}(y,x^t_P)=\{ x^t_Q\in \Xc^t_Q:g_i(y,x^t_{A_i})=x^t_i, \forall i\in [n] \}
\end{align*}

Then, we have 
\begin{align}
&\sum_{y,x^t_P} \sum_{x^t_Q\in G_{\Xc^t_Q}(y,x^t_P)} P_{Y,X^t_{[n]}}(y,x^t_P,x^t_Q)  \nonumber  \\
&=1-{\rm P}_{\rm e}\ge 1-\e.  \label{eq:thm:general:proof:correct:probability}
\end{align}

We also have 
\begin{align}
|G_{\Xc^t_Q}(y,x^t_P)|\le \alpha(\Gamma_t(Q)),  \label{eq:thm:general:proof:alpha}
\end{align}
which can be shown by contradiction as follows. 
%Assume there exists two different realizations $x_Q,x_Q'\in \Xc_Q^{\rm good}(y,x_P)$ such that their corresponding vertices in $\Gamma_t(Q)$ are adjacent. 
Assume there exists two different $x^t_{[n]},z^t_{[n]}\in \Xc^t_{[n]}$, such that $x^t_P=z^t_P$, $x^t_Q\in G_{\Xc^t_Q}(y,x^t_P)$, $z^t_Q\in G_{\Xc^t_Q}(y,x^t_P)$, and $x^t_Q$ and $z^t_Q$ are adjacent (i.e., confusable) in $\Gamma_t(Q)$. 
%In other words, $x^t_Q$ and $z^t_Q$ are confusable in $\Gamma_t(Q)$. 
Hence, there exists some receiver $i\in Q$ such that $x^t_i\neq z^t_i$ and $x^t_{A_i\cap Q}=z^t_{A_i\cap Q}$. 
%With the same $x_P$, we will have $x_i\neq x'_i$ and $x_{A_i}=(x_{A_i\cap P},x_{A_i\cap Q})=(x_{A_i\cap P},x'_{A_i\cap Q})=$
Then considering $x^t_{[n]}$ and $z^t_{[n]}$, since they have the same realizations for messages in $P$, we have $x^t_{A_i}=(x^t_{A_i\cap P},x^t_{A_i\cap Q})=(z^t_{A_i\cap P},z^t_{A_i\cap Q})=z^t_{A_i}$. 
%Therefore, from the perspective of receiver $i$, upon receiving codeword $y$ and observing side information $x^1_{A_i}=x^2_{A_i}$, there are two possible values $x^1_i\neq x^2_i$ for its requested message $i$. 
From the perspective of receiver $i$, upon receiving codeword $y$ and observing side information $x^t_{A_i}=z^t_{A_i}$, it cannot tell whether the true sequence for message $i$ is $x^t_i$ or $z^t_i$. 
Therefore, with the transmitted codeword being $y$, either $x^t_{[n]}$ or $z^t_{[n]}$ being the true realization will lead to an erroneous decoding at receiver $i$, 
which contradicts the assumption that both $x^t_Q$ and $z^t_Q$ belong to $G_{\Xc^t_Q}(y,x^t_P)$. 
%for at least one tuple among $x^1_{[n]}$ and $x^2_{[n]}$
Therefore, any realizations $x^t_Q,z^t_Q\in G_{\Xc^t_Q}(y,x^t_P)$ must be not confusable and thus not adjacent to each other in $\Gamma_t(Q)$. In other words, the vertex subset $G_{\Xc^t_Q}(y,x^t_P)\subseteq V(\Gamma_t(Q))$ must be an independent set in $\Gamma_t(Q)$ and thus its cardinality is upper bounded by the independence number of $\Gamma_t(Q)$.

%\Royy{Set $c(t)^-=\min \{ c(t),|G_{\Xc^t_Q}(y,x^t_P)| \}$}. 
We lower bound ${\rm P_{\rm s}}(X^t_P,Y)$, i.e., the adversary's expected successful guessing probability after observing $Y$, as 
\begin{align}
&\sum_{y,x^t_P} \max_{\substack{K\subseteq \Xc^t_Q:|K|\le c(t)}}\sum_{x^t_Q\in K}P_{Y,X^t_{[n]}}(y,x^t_{[n]})  \nonumber  \\
&\ge \sum_{y,x^t_P}\max_{\substack{K\subseteq G_{\Xc^t_Q}(y,x^t_P):|K|\le c(t)}}\sum_{x^t_Q\in K}P_{Y,X^t_{[n]}}(y,x^t_{[n]})  \nonumber  \\
&\ge \sum_{y,x^t_P}\frac{1}{|\substack{K\subseteq G_{\Xc^t_Q}(y,x^t_P):|K|=c(t)^-}|}  \nonumber  \\
&\qquad \sum_{\substack{K\subseteq G_{\Xc^t_Q}(y,x^t_P):|K|=c(t)^-}}\sum_{x^t_Q\in K}P_{Y,X^t_{[n]}}(y,x^t_{[n]})  \nonumber  \\
&\stackrel{(a)}{=} \sum_{y,x^t_P} \frac{\binom{|G_{\Xc^t_Q}(y,x^t_P)-1|}{c(t)^--1} \sum\limits_{x^t_Q\in G_{\Xc^t_Q}(y,x^t_P)} P_{Y,X^t_{[n]}}(y,x^t_{[n]})}{\binom{|G_{\Xc^t_Q}(y,x^t_P)|}{c(t)^-}}  \nonumber  \\
&=\sum_{y,x^t_P} \frac{c(t)^-}{|G_{\Xc^t_Q}(y,x^t_P)|} \sum_{x^t_Q\in G_{\Xc^t_Q}(y,x^t_P)} P_{Y,X^t_{[n]}}(y,x^t_{[n]})  \nonumber  \\
&\stackrel{(b)}{\ge} \frac{c(t)(1-\epsilon)}{\alpha(\Gamma_t(Q))},  \label{eq:index:coding:Ae:multi:guess:numerator:lower:ddd}
%&\stackrel{(b)}{\ge} \frac{c(t)^-}{|G_{\Xc^t_Q}(y,x^t_P)|} \cdot (1-\e) \stackrel{(c)}{\ge} \frac{c(t)(1-\epsilon)}{\alpha(\Gamma_t(Q))},  \label{eq:index:coding:Ae:multi:guess:numerator:lower:ddd}
\end{align}
where 
$c(t)^-=\min \{ c(t),|G_{\Xc^t_Q}(y,x^t_P)| \}$, and 
%
%and $\Xc^t_{A_e^c}(y,x^t_{A_e},{\rm good})$ denotes the collection of $x^t_{A_e^c}$ such that $(y,x^t_{A_e},x^t_{A_e^c})$ are jointly possible and can be correctly decoded at every receiver. 
%The last inequality above follows from the fact that $|\Xc^t_{A_e^c}(y,x^t_{A_e},{\rm good})|\le \alpha(\Gamma_t(A_e^c))$ and that $$\sum_{y}\sum_{x^t_{A_e}}\sum_{x^t_{A_e^c}\in \Xc^t_{A_e^c}(y,x^t_{A_e},{\rm good})} P_{Y,X^t_{[n]}}(y,x^t_{[n]})=1-P_e\ge 1-\epsilon. $$
\begin{itemize}
\item (a) follows from the fact that each $x^t_Q\in G_{\Xc^t_Q}(y,x^t_P)$ appears in exactly $\binom{|G_{\Xc^t_Q}(y,x^t_P)-1|}{c(t)^{-}-1}$ subsets of $G_{\Xc^t_Q}(y,x^t_P)$ of size $c(t)^-$, 
\item (b) follows from \eqref{eq:thm:general:proof:correct:probability}, \eqref{eq:thm:general:proof:alpha}, and that if $c(t)\le |G_{\Xc^t_Q}(y,x^t_P)|$, then $\frac{c(t)^-}{|G_{\Xc^t_Q}(y,x^t_P)|}=\frac{c(t)}{|G_{\Xc^t_Q}(y,x^t_P)|}\ge \frac{c(t)}{\alpha(\Gamma_t(Q))}$, otherwise we have $c(t)>|G_{\Xc^t_Q}(y,x^t_P)|$ and thus $\frac{c(t)^-}{|G_{\Xc^t_Q}(y,x^t_P)|}=1\ge \frac{c(t)}{\alpha(\Gamma_t(Q))}$, where the inequality is due to the assumption that $c(t)\le \alpha(\Gamma_t(Q))$.
\iffalse
\item (b) follows from \eqref{eq:thm:general:proof:correct:probability}, 
%that $\sum_{y,x^t_P} \sum_{x^t_Q\in \Xc^t_Q(y,x^t_P,{\rm good})} P_{Y,X^t_{[n]}}(y,x^t_{[n]})=1-{\rm P}_{\rm e}\ge 1-\e$, 
\item (c) follows from \eqref{eq:thm:general:proof:alpha} and thus, if $c(t)\le |G_{\Xc^t_Q}(y,x^t_P)|$, then $\frac{c(t)^-}{|G_{\Xc^t_Q}(y,x^t_P)|}=\frac{c(t)}{|G_{\Xc^t_Q}(y,x^t_P)|}\ge \frac{c(t)}{\alpha(\Gamma_t(Q))}$, otherwise we have $c(t)>|G_{\Xc^t_Q}(y,x^t_P)|$ and thus $\frac{c(t)^-}{|G_{\Xc^t_Q}(y,x^t_P)|}=1\ge \frac{c(t)}{\alpha(\Gamma_t(Q))}$, where the inequality is due to the assumption that $c(t)\le \alpha(\Gamma_t(Q))$. 
\fi
\end{itemize}

%\Royy{we need to use $x^t_{[n]}$ rather than $x_{[n]}$}

%Recall that \eqref{eq:index:coding:Ae:multi:guess:numerator:lower:ddd} gives 
%\begin{align}
%\sum_{y}\sum_{x^t_{A_e}}\max_{\substack{K\subseteq \Xc^t_{A_e^c}:\\|K|\le g(t)}}\sum_{x^t_{A_e^c}\in K}P_{Y,X^t_{[n]}}
%%&\ge \sum_{y}\sum_{x^t_{A_e}} \frac{g(t)^-}{|\Xc^t_{A_e^c}(y,x^t_{A_e},{\rm good})|} \sum_{x^t_{A_e^c}\in \Xc^t_{A_e^c}(y,x^t_{A_e},{\rm good})} P_{Y,X^t_{[n]}}(y,x^t_{[n]})  \nonumber  \\
%%%
%&\ge \frac{g(t)(1-\epsilon)}{\alpha(\Gamma_t(A_e^c))}. \label{eq:index:coding:Ae:multi:guess:general:vanishing:numerator:lower}
%\end{align}

%As for the denominator in \eqref{eq:def:leakage:rate:any:t:f}, ${\rm P_{\rm s}}(X^t_P)$, 
For bounding ${\rm P_{\rm s}}(X^t_P)$, 
consider any two disjoint subsets $A,B\subseteq [n]$. %and any message length $t$. 
%Denote $x_A$ as $(x_{A,1},x_{A,2},\ldots,x_{A,t})$ and any $x_B=(x_{B,1},x_{B,2},\ldots,x_{B,t})$. 
Note that any realization $x^t_A$ can be explicitly denoted as $(x_{A,1},x_{A,2},\ldots,x_{A,t})$. 
We have 
\begin{align}
&\sum_{x^t_A} \max_{x^t_B} P_{X^t_{A\cup B}}(x^t_A,x^t_B)  \nonumber  \\
&=\sum_{x^t_A} \max_{x^t_B} \prod_{j\in [t]}P_{X_{A\cup B}}(x_{A,j},x_{B,j})  \nonumber  \\
%%
%&=\sum_{x^t_A} \prod_{j\in [t]}P_{X_{A\cup B}}(x_{A,j},x_{B,j}(x_{A,j}))  \nonumber  \\
%%
&=\big( \sum_{x_A} \max_{x_B} P_{X_{A\cup B}}(x_A,x_B) \big)^t, \label{eq:index:coding:Ae:multi:guess:general:vanishing:denominator}
\end{align}
%\begin{align}
%\sum_{x^t_A} \max_{x^t_B} P_{X^t_{A\cup B}}(x^t_A,x^t_B)&=\sum_{x^t_A} \max_{x^t_B} \prod_{j\in [t]}P_{X_{A\cup B}}(x_{A,j},x_{B,j})  \nonumber  \\
%%%
%%&=\sum_{x^t_A} \prod_{j\in [t]}P_{X_{A\cup B}}(x_{A,j},x_{B,j}(x_{A,j}))  \nonumber  \\
%%%
%&=\big( \sum_{x_A} \max_{x_B} P_{X_{A\cup B}}(x_A,x_B) \big)^t, \label{eq:index:coding:Ae:multi:guess:general:vanishing:denominator}
%\end{align}
where the last equality can be shown via induction. %by mathematical induction. 

Based on \eqref{eq:index:coding:Ae:multi:guess:numerator:lower:ddd} and \eqref{eq:index:coding:Ae:multi:guess:general:vanishing:denominator}, we have 
%Based on \eqref{eq:index:coding:Ae:multi:guess:general:vanishing:numerator:lower} and \eqref{eq:index:coding:Ae:multi:guess:general:vanishing:denominator}, we have 
\begin{align}
\Lc&\ge \lim_{\e\to 0}\lim_{t\to \infty} \frac{1}{t} \log \frac{\frac{c(t)(1-\epsilon)}{\alpha(\Gamma_t(Q))}}{ \sum_{x^t_P} \max\limits_{K\subseteq \Xc^t_Q:|K|\le c(t)} \sum\limits_{x^t_Q\in K} P_{X^t_{[n]}}(x^t_{[n]}) }  \nonumber  \\
&\ge \lim_{\e\to 0}\lim_{t\to \infty} \frac{1}{t} \log \frac{\frac{c(t)(1-\epsilon)}{\alpha(\Gamma_t(Q))}}{c(t)\cdot \sum_{x^t_P} \max_{x^t_Q} P_{X^t_{[n]}}(x^t_{[n]})}  \nonumber  \\
&=\lim_{\e\to 0}\lim_{t\to \infty} \frac{1}{t} \log \frac{\frac{(1-\e)|V(\Gamma_t(Q))|}{\alpha(\Gamma_t(Q))} \cdot \frac{1}{|V(\Gamma_t(Q))|}}{(\sum_{x_P}\max_{x_Q} P_{X_{[n]}}(x_{[n]}))^t}  \nonumber  \\
&\stackrel{(c)}{=} \lim_{t\to \infty} \frac{1}{t} \log \chi_{\rm f}(\Gamma_t(Q)) + \log \frac{|\Xc|^{-|Q|}}{\sum_{x_P}\max_{x_Q} P_{X_{[n]}}(x_{[n]})}  \nonumber  \\
%&=\lim_{t\to \infty} \frac{1}{t} \log \chi_f(\Gamma_t(A_e^c)) - |A_e^c| + \log \frac{1}{\sum_{x_{A_e}} \max_{x_{A_e^c}} P_{X_{[n]}}(x_{[n]})}  \nonumber  \\
%%%
&\stackrel{(d)}{=} \rho(Q) - |Q| + \log \frac{1}{\sum_{x_P}\max_{x_Q} P_{X_{[n]}}(x_{[n]})},  \nonumber
\end{align}
where (c) follows from the fact that for any vertex-transitive graph $G$, $\chi_{\rm f}(G)=|V(G)|/\alpha(G)$ \cite[Proposition 3.1.1]{scheinerman2011fractional}, and that any confusion graph for index coding is vertex-transitive \cite[Section 11.4]{arbabjolfaei2018fundamentals}, 
and (d) follows from \eqref{eq:compression:zero:error}. 
%
%which completes the proof of the lower bound in \eqref{eq:index:coding:Ae:multi:guess:general:vanishing}. 
\end{IEEEproof}

\begin{IEEEproof}[Proof of the upper bound in \eqref{eq:index:coding:Ae:multi:guess:general:vanishing}]
%Now we show the upper bound in \eqref{eq:index:coding:Ae:multi:guess:general:vanishing}. Towards that end, 
%Consider any $(t,M,f,\gv)$ index code that maps messages $X^t_{[n]}$ to codeword $Y=(Y_1,Y_2)$ according to the following rules. 
Consider any decoding error $\e>0$. 
Construct a deterministic encoding function $f$ that maps messages $X^t_{[n]}$ to codeword $Y=(Y_1,Y_2)$ according to the following rules. 
\begin{enumerate}
\item Codeword $Y_1$ is generated from $X^t_P$ according to some deterministic encoding function $f_1:\Xc^{t|P|}\to \{1,2,\ldots,|\Yc_1|\}$ such that there exist some decoding functions $g_i,i\in P$ allowing zero-error decoding for all receivers $i\in P$ and that $t^{-1}\log |\Yc_1|=\rho(P)$. 
\item Codeword $Y_2$ is generated from $X^t_Q$ according to some deterministic encoding function $f_2:\Xc^{t|Q|}\to \{1,2,\ldots,|\Yc_2|\}$ such that there exist some decoding functions $g_i,i\in Q$ allowing $\e$-error decoding for all receivers $i\in Q$ and that $t^{-1}\log |\Yc_2|=\Rc(Q)$. 
\end{enumerate}
%to codeword $Y$ satisfying the following conditions: $Y=(Y_1,Y_2)$ where $Y_1$ is generated from $X^t_{A_e}$ according to some deterministic zero-error mapping scheme for the subproblem $\Gamma(A_e)$ achieving the zero-error broadcast rate $\beta(\Gamma(A_e),P_{X_{A_e}})$, and $Y_2$ is generated from $X^t_{A_e^c}$ according to some deterministic vanishing-error mapping scheme for the subproblem $\Gamma(A_e^c)$ achieving the vanishing-error broadcast rate $\beta^*(\Gamma(A_e),P_{X_{A_e}})$. 
%%
Such encoding functions $f_1$ and $f_2$ exist for sufficiently large $t$. 
%We further verify that the composite $(t,|\Yc_1|\cdot |\Yc_2|,f,\gv)$ index code described above allows $\e$-error decoding and thus is valid. 
We further verify that the coding scheme described above leads to an average probability of error ${\rm P}_{\rm e}$ no more than $\e$ and thus is valid. 
%Such mapping scheme $P_{Y|X^t_{[n]}}$ exists for sufficiently large $t$ and can be shown to be valid for the problem $\Gamma$. 
%\Royy{definition of the code not very clear, check 1.A in index coding monograph}
%Consider any decoding error $\e>0$. 
%Let $\Xc^t_Q({\rm bad})$ denote the collection of realizations $x^t_Q$ such that 
Note that $f_1$ and $f_2$ are all deterministic. 
%Define $\Xc^t_Q({\rm bad})\doteq \{ x^t_Q: \text{there exists some $i\in Q$ such that $g_i(f_2(x^t_Q))\neq x^t_i$} \}. $ 
Define $B_{\Xc^t_Q}=\{ x^t_Q: \text{there exists some $i\in Q$ such that $g_i(f_2(x^t_Q))\neq x^t_i$} \}. $ 
That is, $B_{\Xc^t_Q}$ denotes the set of $x^t_Q$ for which there is at least one receiver $i\in Q$ that decodes erroneously. 
We have 
\begin{align}
\e\ge \sum_{x^t_Q\in B_{\Xc^t_Q}} P_{X^t_Q}(x^t_Q).  \label{eq:thm:general:upper:error}
\end{align}
%
%Therefore, for the $(t,M,f,\gv)$ index code, we have
Hence, we have 
\begin{align*}
{\rm P}_{\rm e}
%&=\sum_{y=(y_1,y_2)}\sum_{x^t_P}\sum_{x^t_Q\in \Xc^t_Q(y,x^t_P,{\rm bad})} P_{Y_1,Y_2,X^t_{[n]}}(y_1,y_2,x^t_{[n]})  \\
&=\sum_{x^t_P} \sum_{x^t_Q\in B_{\Xc^t_Q}} P_{X^t_{[n]}}(x^t_{[n]})  \\
&=\sum_{x^t_Q\in B_{\Xc^t_Q}} P_{X^t_Q}(x^t_Q) \sum_{x^t_P} P_{X^t_P|X^t_Q}(x^t_P|x^t_Q)  \\
&=\sum_{x^t_Q\in B_{\Xc^t_Q}} P_{X^t_Q}(x^t_Q) \cdot 1\le \e,
\end{align*}
where the last inequality follows from \eqref{eq:thm:general:upper:error}. 
Now we have shown that the proposed coding scheme is valid. 

The optimal leakage rate is upper bounded by the rate of the information leakage of the proposed coding scheme as $\e$ goes to $0$. 
Let $P_{Y,X^t_{[n]}}$ denote the joint distribution of $Y=(Y_1,Y_2)$ and $X^t_{[n]}$ according to the proposed coding scheme. 
For any $x^t_P\in \Xc^t_P$ and $y_2\in \Yc_2$, define
\begin{align}
\Xc^t_Q(x^t_P,y_2)&=\{ x^t_Q\in \Xc^t_Q: P_{Y_2,X^t_{[n]}}(y_2,x^t_P,x^t_Q)>0 \},  \nonumber%\label{eq:thm:general:upper:subset}  %\\
%\Xc^t_Q(x^t_P)&=\{ x^t_Q\in \Xc^t_Q: P_{X^t_{[n]}}(x^t_P,x^t_Q)>0 \}.
\end{align}
%Clearly, $\Xc^t_Q(x^t_P,y_2)\subseteq \Xc^t_Q$. 
%For any $x^t_P\in \Xc^t_P$ and $y_2\in \Yc_2$, let $\Xc^t_Q(x^t_P,y_2)$ denote the collection of $x^t_Q$ that is jointly possible with $x^t_P$ and $y_2$, and let $\Xc^t_Q(x^t_P)$ denote the collection of $x^t_Q$ that is jointly possible with $x^t_P$. 
%Clearly, $\Xc^t_Q(x^t_P,y_2)\subseteq \Xc^t_Q(x^t_P)$. 
Then we have 
%\Royy{do we $\e\to 0$ in the derivation?}
%With such mapping scheme, we have 
\begin{align}
\Lc
%&\Lc  \nonumber  \\
%%%
%&=\lim_{t\to \infty} \frac{1}{t} \log \frac{\sum_{y}\sum_{x^t_{A_e}} \max_{\substack{K\subseteq \Xc^t_{A_e^c}:|K|\le g(t)}}\sum_{x^t_{A_e^c}\in K}P_{Y,X^t_{[n]}}(y,x^t_{[n]})}{\sum_{x^t_{A_e}} \max_{\substack{K\subseteq \Xc^t_{A_e^c}:\\|K|\le g(t)}}\sum_{x^t_{A_e^c}\in K}P_{X^t_{[n]}}(x^t_{[n]})}  \nonumber  \\
%%
&\le \lim_{\e\to 0}\lim_{t\to \infty} \frac{1}{t} \log \frac{   \sum\limits_{\substack{x^t_P,\\y_1,y_2}} \max\limits_{\substack{K\subseteq \Xc^t_Q:\\|K|\le c(t)}} \sum\limits_{x^t_Q\in K} P_{Y,X^t_{[n]}}(y_1,y_2,x^t_{[n]})   }{\sum\limits_{x^t_P} \max\limits_{\substack{K\subseteq \Xc^t_Q:\\|K|\le c(t)}}\sum_{x^t_Q\in K}P_{X^t_{[n]}}(x^t_{[n]})}  \nonumber  \\
&\stackrel{(a)}{=} \lim_{\e\to 0}\lim_{t\to \infty} \frac{1}{t} \log \frac{   \sum\limits_{x^t_P,y_2} \max\limits_{\substack{K\subseteq \Xc^t_Q(x^t_P,y_2):\\|K|\le c(t)}} \sum\limits_{x^t_Q\in K} P_{X^t_{[n]}}(x^t_{[n]})   }{\sum\limits_{x^t_P} \max\limits_{\substack{K\subseteq \Xc^t_Q:\\|K|\le c(t)}}\sum\limits_{x^t_Q\in K}P_{X^t_{[n]}}(x^t_{[n]})}  \nonumber  \\
&\stackrel{(b)}{\le} \lim_{\e\to 0}\lim_{t\to \infty} \frac{1}{t} \log \frac{|\Yc_2|\cdot (\sum\limits_{x^t_P} \max\limits_{\substack{K\subseteq \Xc^t_Q:\\|K|\le c(t)}}\sum\limits_{x^t_Q\in K}P_{X^t_{[n]}}(x^t_{[n]}))}{\sum\limits_{x^t_P} \max\limits_{\substack{K\subseteq \Xc^t_Q:\\|K|\le c(t)}}\sum\limits_{x^t_Q\in K}P_{X^t_{[n]}}(x^t_{[n]})}  \nonumber  \\
&=\lim_{\e\to 0}\lim_{t\to \infty} \frac{1}{t} \log |\Yc_2| = \Rc(Q),  \nonumber
\end{align}
where (a) follows from the definition of $\Xc^t_Q(x^t_P,y_2)$ %in \eqref{eq:thm:general:upper:subset} 
and the fact that $Y_1$ is a deterministic function of $X^t_P$ and $Y_2$ is a deterministic function of $X^t_Q$, 
and (b) follows from $\Xc^t_Q(x^t_P,y_2)\subseteq \Xc^t_Q$. 
%and (c) follows from the construction of the code. 
%
%which completes the proof of the upper bound in \eqref{eq:index:coding:Ae:multi:guess:general:vanishing}. 
\iffalse
The lower bound in \eqref{eq:index:coding:Ae:multi:guess:general:zero:error} follows directly from the lower bound in \eqref{eq:index:coding:Ae:multi:guess:general:vanishing} and the fact that $\Lc\le \lambda$. 
%It remains to show the upper bound 
The upper bound in \eqref{eq:index:coding:Ae:multi:guess:general:zero:error} can be shown using similar techniques to the proof of the upper bound in \eqref{eq:index:coding:Ae:multi:guess:general:vanishing}. 
\fi
\end{IEEEproof}

\begin{remark}
%An observation is that Theorem \ref{thm:general} is independent of the guessing capability function $c(t)$. 
\Royy{An interesting observation is that the bounds in Theorem \ref{thm:general} is independent of the guessing capability function $c(t)$. 
Whether $\Lc$ and $\lambda$ depend on $c(t)$ remains unclear. 
}
\end{remark}

%\begin{remark}
%The coding scheme proposed in the achievability proof of Theorem \ref{thm:general} achieves broadcast rate $\Rc$ iff $\Rc=\rho(P)+\Rc(Q)$. 
%Similarly, for the zero-error scenario, the coding scheme BLABLA
%\end{remark}

The upper and lower bounds in Theorem \ref{thm:general} do not match in general, as shown in the following example.

%\begin{example}
Consider the $4$-message index coding problem $(1|4),(2|3),(3|2),(4|1)$, where the messages are binary and independent of each other with $P_{X_1}(0)=1/4$ and $P_{X_1}(1)=3/4$, and $X_2$, $X_3$, and $X_4$ all follow a uniform distribution. Consider an adversary knowing $X_P=X_4$ as side information, and thus $Q=\{1,2,3\}$. 
%Note that the induced subproblem by $Q$ is the $3$-message problem $(1|-),(2|3),(3|2)$ with joint distribution $P_{X_{\{1,2,3\}}}$, which has been studied in \cite{miyake2015index}. 
%For this problem, Theorem \ref{thm:general} gives 
%\begin{align*}
%\Rc(Q)&\stackrel{(a)}{=}H(X_1)+\max \{ H(X_2|X_{1,3}),H(X_3|X_{1,2}) \}  \\
%%&=\frac{1}{4}\log 4+3/4\cdot \log \frac{4}{3}+1  \\
%&=3-\frac{3}{4}\log 3,
%\end{align*}
%as an upper bound on the leakage rate $\Lc$, where (a) has been shown in \cite[Proposition 1]{miyake2015index}. 
The broadcast rate for the subproblem induced by $Q$ has been previously found \cite{miyake2015index} to be $\Rc(Q)=3-\frac{3}{4}\log 3$. 
By Theorem \ref{thm:general}, the leakage rate $\Lc$ is upper bounded by $\Rc(Q)$, and lower bounded as 
%And, Theorem \ref{thm:general} gives 
\begin{align*}
\Lc\ge &\rho(Q)-|Q|+\log \frac{1}{\sum_{x_{P}}\max_{x_{Q}}P_{X_{[n]}}(x_{[n]})}  \\
&=2-3+\log \frac{1}{3/32+3/32}=3-\log 3. 
\end{align*}
%as a lower bound on $\Lc$. 
Note that $\rho(Q)=2$ can be easily verified (for example, see  \cite[Section 8.6]{arbabjolfaei2018fundamentals}). 
For the zero-error leakage rate $\lambda$, by \eqref{eq:index:coding:Ae:multi:guess:general:zero:error} in Theorem \ref{thm:general}, we have
\begin{align*}
3-\log 3\le \lambda\le 2.
\end{align*}
\iffalse
Consider the $4$-message index coding problem $(1|4),(2|3),(3|2),(4|1)$, where the messages are binary and independent of each other with $P_{X_1}(0)=1/4$ and $P_{X_1}(1)=3/4$, and $X_2$, $X_3$, and $X_4$ all following a uniform distribution. 
Consider an adversary knowing $X_P=X_4$ as side information, and thus $Q=\{1,2,3\}$. 
Theorem \ref{thm:general} gives 
\begin{align*}
&\rho(Q)-|Q|+\log \frac{1}{\sum_{x_{P}}\max_{x_{Q}}P_{X_{[n]}}(x_{[n]})}  \\
&=2-3+\log \frac{1}{3/32+3/32}=3-\log 3
\end{align*}
as a lower bound on the leakage rate $\Lc$, where $\rho(Q)=2$ is a known result in \cite[Section 8.6]{arbabjolfaei2018fundamentals}, 
and 
\begin{align*}
\Rc(Q)&\stackrel{(a)}{=}H(X_1)+\max \{ H(X_2|X_{1,3}),H(X_3|X_{1,2}) \}  \\
%&=\frac{1}{4}\log 4+3/4\cdot \log \frac{4}{3}+1  \\
&=3-\frac{3}{4}\log 3,
\end{align*}
as upper bound on $\Lc$, where (a) follows from \cite[Proposition 1]{miyake2015index}. 
%It remains clear whether the lower or the upper bound is tight for this problem, or the actual leakage rate 
The exact value of $\Lc$ for this problem remains unclear. 
\fi
%\end{example}

%In the next two subsections, we consider two special cases with extra assumptions on the message distribution $P_{X_{[n]}}$, where the lower and upper bounds in Theorem \ref{thm:general} match and thus establish the optimal rate. 

%\Royy{the next two sections to be rewritten}

%%%%%%%%%%%%%%%%%%%%%%%%%%%%%%%%%%%%%%%%%%%%%
%%---------------------
%%---------------------
%%---------------------
%%%%%%%%%%%%%%%%%%%%%%%%%%%%%%%%%%%%%%%%%%%%%

\subsection{Leakage Under A Uniform Message Distribution}

In most existing works for index coding, the messages $X_{[n]}$ are assumed to be uniformly distributed and thus independent of each other. 
In such cases, Theorem \ref{thm:general} simplifies to the following corollary. 

\iffalse
With uniformly distributed independent sources, it has been shown in \cite{Langberg--Effros2011} that the zero-error compression rate is equal to the vanishing-error compression rate. 
%Thus in this subsection we may refer to both the zero-error and vanishing-error compression rate simply as the compression rate. 

The zero-error compression rate $\beta(\Gamma)$ and vanishing-error compression rate $\beta^*(\Gamma)$ are equal and can be expressed based on the confusion graph $\Gamma$ as 
\begin{align}
\beta(\Gamma)=\beta^*(\Gamma)=\lim_{t\to \infty}\frac{1}{t}\log \chi(\Gamma_t)\stackrel{(a)}{=}\lim_{t\to \infty}\frac{1}{t}\log \chi(\Gamma_t),  \label{eq:index:coding:compression:uniform}
\end{align}
where $(a)$ has been shown in \cite[Section 3.2]{arbabjolfaei2018fundamentals}.

We consider the general adversary model where the adversary knows some messages as side information and can make multiple guesses after observation of each codeword. 
More specifically, let $A_e$ denote the adversary's side information, and the adversary is interested in guessing its unknown messages $X_{A_e^c}$. 
As for the adversary guessing capability function $g(t)$, we make similar assumptions as we did in \cite{liu2021information} (particularly, $g(t)\le \alpha(\Gamma_t(A_e^c))$ where $\alpha(\cdot)$ denotes the independence number of a graph and $\Gamma_t(A_e^c)$ denotes the confusion graph of the subproblem induced by set $A_e^c$ with message length being $t$. 
\fi

\begin{corollary}  \label{cor:uniform}
%Consider any problem $\Gamma$. Let $\Lc_{g,A_e}$ and $\Lc_{g,A_e}^*$ denote the zero-error and vanishing-error leakage rate, respectively. 
If $P_{X_{[n]}}$ follows a uniform distribution, then  
\begin{align}
\Lc=\lambda=\Rc(Q)=\rho(Q).  \label{eq:cor:uniform}
\end{align}
\end{corollary}
%\lim_{t\to \infty}\frac{1}{t}\log \chi_f(\Gamma_t(A_e^c))

\begin{IEEEproof}
We have
\begin{align*}
&\rho(Q)-|Q|+\log \frac{1}{\sum_{x_{P}}\max_{x_{Q}}P_{X_{[n]}}(x_{[n]})}  \nonumber  \\
&=\rho(Q)-|Q|+\log \frac{1}{|\Xc|^{t|P|}\cdot (1/|\Xc|^{tn})}  \nonumber  \\
&=\rho(Q)=\Rc(Q),
\end{align*}
where the last equality follows from the fact that the vanishing-error and zero-error broadcast rates are equal when messages are uniformly distributed \cite{Langberg--Effros2011}. 
Combining Theorem \ref{thm:general} and the above result yields \eqref{eq:cor:uniform}. 
\end{IEEEproof}

\begin{remark}
%As there is no known single-letter expression for the broadcast rate of index coding even under uniform message distribution, the characterization of leakage rates in Corollary \ref{cor:uniform} is not single-letter either. 
Even though we have established the equivalence between the leakage and broadcast rates under uniform message distribution, a computable single-letter characterization of the value in \eqref{eq:cor:uniform} is unknown. 
%the value is in general not computable
%
Nevertheless, the equivalence between the leakage and broadcast rates means that the extensive results on the broadcast rate of index coding established in the literature (such as single-letter lower and upper bounds, explicit characterization for special cases, and structural properties) can be directly used to determine or bound the leakage rate. 
%the equivalence between $\Lc_{g,A_e}$ (or $\Lc_{g,A_e}^*$) and $\beta(\Gamma(A_e^c))$ is of significance. This is because that the compression rate $\beta$ has been extensively studied in the literature with many results established (lower and upper bounds, exact value for special cases, structural properties and cardinalities, etc.) and all these results can be used to determine or bound the leakage rate $\Lc_{g,A_e}$ due to the equivalence. 
\end{remark}

\begin{remark}
As the leakage rate in \eqref{eq:cor:uniform} can be achieved by the proposed coding scheme in the achievability proof of Theorem \ref{thm:general}, for any index coding instance with uniform message distribution satisfying $\Rc=\Rc(P)+\Rc(Q)$ (or equivalently, $\rho=\rho(P)+\rho(Q)$), we know that the broadcast rate and leakage rate can be simultaneously achieved by some deterministic index code. 
\end{remark}

\bibliographystyle{IEEEtran}
\bibliography{references}

% Generated by IEEEtran.bst, version: 1.14 (2015/08/26)
\newcommand{\noopsort}[1]{}
\begin{thebibliography}{10}
\providecommand{\url}[1]{#1}
\csname url@samestyle\endcsname
\providecommand{\newblock}{\relax}
\providecommand{\bibinfo}[2]{#2}
\providecommand{\BIBentrySTDinterwordspacing}{\spaceskip=0pt\relax}
\providecommand{\BIBentryALTinterwordstretchfactor}{4}
\providecommand{\BIBentryALTinterwordspacing}{\spaceskip=\fontdimen2\font plus
\BIBentryALTinterwordstretchfactor\fontdimen3\font minus
  \fontdimen4\font\relax}
\providecommand{\BIBforeignlanguage}[2]{{%
\expandafter\ifx\csname l@#1\endcsname\relax
\typeout{** WARNING: IEEEtran.bst: No hyphenation pattern has been}%
\typeout{** loaded for the language `#1'. Using the pattern for}%
\typeout{** the default language instead.}%
\else
\language=\csname l@#1\endcsname
\fi
#2}}
\providecommand{\BIBdecl}{\relax}
\BIBdecl

\bibitem{Birk--Kol1998}
Y.~Birk and T.~Kol, ``Informed-source coding-on-demand ({ISCOD}) over broadcast
  channels,'' in \emph{IEEE INFOCOM}, Mar. 1998, pp. 1257--1264.

\bibitem{bar2011index}
Z.~Bar-Yossef, Y.~Birk, T.~Jayram, and T.~Kol, ``Index coding with side
  information,'' \emph{{IEEE} Trans. Inf. Theory}, vol.~57, pp. 1479--1494,
  2011.

\bibitem{arbabjolfaei2018fundamentals}
F.~Arbabjolfaei and Y.-H. Kim, ``Fundamentals of index coding,''
  \emph{Foundations and Trends{\textregistered} in Communications and
  Information Theory}, vol.~14, no. 3-4, pp. 163--346, 2018.

\bibitem{dau2012security}
S.~H. Dau, V.~Skachek, and Y.~M. Chee, ``On the security of index coding with
  side information,'' \emph{{IEEE} Trans. Inf. Theory}, vol.~58, no.~6, pp.
  3975--3988, 2012.

\bibitem{ong2016secure}
L.~Ong, B.~N. Vellambi, P.~L. Yeoh, J.~Kliewer, and J.~Yuan, ``Secure index
  coding: Existence and construction,'' in \emph{Proc. {IEEE} Int. Symp. on
  Information Theory ({ISIT})}, Barcelona, Spain, 2016, pp. 2834--2838.

\bibitem{liu:vellambi:kim:sadeghi:itw18}
Y.~Liu, Y.-H. Kim, B.~Vellambi, and P.~Sadeghi, ``On the capacity region for
  secure index coding,'' in \emph{Proc. {IEEE} Information Theory Workshop
  ({ITW})}, Guanzhou, China, Nov. 2018.

\bibitem{ong2021code}
L.~Ong, B.~N. Vellambi, J.~Kliewer, and P.~L. Yeoh, ``A code and rate
  equivalence between secure network and index coding,'' \emph{IEEE J. Sel.
  Areas Inf. Theory}, vol.~2, no.~1, pp. 106--120, 2021.

\bibitem{liu2020secure}
Y.~Liu, P.~Sadeghi, N.~Aboutorab, and A.~Sharififar, ``Secure index coding with
  security constraints on receivers,'' in \emph{Proc. {IEEE} Int. Symp. on
  Information Theory and its Applications ({ISITA})}, Oct. 2020.

\bibitem{private:index:coding:arxiv}
V.~Narayanan, J.~Ravi, V.~K. Mishra, B.~K. Dey, N.~Karamchandani, and V.~M.
  Prabhakaran, ``Private index coding,'' \emph{arXiv preprint
  arXiv:2006.00257}, 2020.

\bibitem{karmoose2019privacy}
M.~Karmoose, L.~Song, M.~Cardone, and C.~Fragouli, ``Privacy in index coding: $
  k $-limited-access schemes,'' \emph{{IEEE} Trans. Inf. Theory}, vol.~66,
  no.~5, pp. 2625--2641, 2019.

\bibitem{yuchengliu2020isit}
Y.~{Liu}, N.~{Ding}, P.~{Sadeghi}, and T.~{Rakotoarivelo}, ``Privacy-utility
  tradeoff in a guessing framework inspired by index coding,'' in \emph{Proc.
  {IEEE} Int. Symp. on Information Theory ({ISIT})}, Jun. 2020, pp. 926--931.

\bibitem{smith2009foundations}
G.~Smith, ``On the foundations of quantitative information flow,'' in
  \emph{International Conference on Foundations of Software Science and
  Computational Structures}.\hskip 1em plus 0.5em minus 0.4em\relax Springer,
  2009, pp. 288--302.

\bibitem{braun2009quantitative}
C.~Braun, K.~Chatzikokolakis, and C.~Palamidessi, ``Quantitative notions of
  leakage for one-try attacks,'' \emph{Electronic Notes in Theoretical Computer
  Science}, vol. 249, pp. 75--91, 2009.

\bibitem{issa2019operational}
I.~Issa, A.~B. Wagner, and S.~Kamath, ``An operational approach to information
  leakage,'' \emph{{IEEE} Trans. Inf. Theory}, vol.~66, no.~3, pp. 1625--1657,
  2019.

\bibitem{isit:2021:arxiv}
Y.~Liu, L.~Ong, S.~Johnson, J.~Kliewer, P.~Sadeghi, and P.~L. Yeoh,
  ``Information leakage in zero-error source coding: A graph-theoretic
  perspective,'' in \emph{Proc. {IEEE} Int. Symp. on Information Theory
  ({ISIT})}, Melbourne, Australia, 2021.

\bibitem{korner1973coding}
J.~K{\"o}rner, ``Coding of an information source having ambiguous alphabet and
  the entropy of graphs,'' in \emph{6th Prague Conference on Information
  Theory}, 1973, pp. 411--425.

\bibitem{alon2008broadcasting}
N.~Alon, E.~Lubetzky, U.~Stav, A.~Weinstein, and A.~Hassidim, ``Broadcasting
  with side information,'' in \emph{49th Annu. IEEE Symp. on Foundations of
  Computer Science (FOCS)}, Oct. 2008, pp. 823--832.

\bibitem{scheinerman2011fractional}
E.~R. Scheinerman and D.~H. Ullman, \emph{Fractional Graph Theory: A Rational
  Approach to the Theory of Graphs}.\hskip 1em plus 0.5em minus 0.4em\relax
  Courier Corporation, 2011.

\bibitem{Langberg--Effros2011}
M.~Langberg and M.~Effros, ``Network coding: {I}s zero error always possible?''
  in \emph{Proc. 49th Ann. Allerton Conf. Comm. Control Comput.}, 2011, pp.
  1478--1485.

\bibitem{miyake2015index}
S.~Miyake and J.~Muramatsu, ``Index coding over correlated sources,'' in
  \emph{Proc. {IEEE} Int. Symp. on Network Coding (NetCod)}, Sydney, Australia,
  2015, pp. 36--40.

\end{thebibliography}

\end{document}